\documentclass[prl,twocolumn,epsf,psfig]{revtex4}
\usepackage{graphicx}
\DeclareGraphicsExtensions{.pdf,.png,.gif,.jpg}
\usepackage{float}
\usepackage{subfig}
\usepackage{epsfig}
\usepackage{wrapfig}

\begin{document} \title{A statistical mechanics perspective for protein folding from $q$-state Potts model}

\author{Theja N. De Silva$^{1,2, 3}$ and Vattika Sivised$^{1}$}
\affiliation{1. Department of Chemistry and Physics,
Augusta University, Augusta, Georgia 30912, USA;\\
2. Kavli Institute for Theoretical Physics, University of California, Santa Barbara, California 93106, USA; \\
 and 3. Institute for Theoretical Atomic, Molecular, and Optical Physics,Harvard-Smithsonian Center for Astrophysics, Harvard University, Cambridge, Massachusetts 02138, USA.}

\begin{abstract}
The folding of a peptide chain into a three dimensional structure is a thermodynamically driven process such that the chain naturally evolves to form domains of similar amino acids. The formation of this domain occurs by curling the one dimensional amino acid sequence by moving similar amino acids proximity to each other. We model this formation of domains or “ordering of amino acids” using q-state Potts model and study the thermodynamic properties using a statistical mechanics approach. Converting the interacting amino acids into an effectively non-interacting model using a mean-field theory, we calculate the Helmholtz free energy (HFE). Then by investigating the HFE, we study the properties of protein folding transition qualitatively. We find that the protein folding phase transition is a strongly first order and the specific heat shows the experimental signatures of this phase transition. Further, we compare these mean-field results with exact transfer matrix results in one dimension and then large $q$ expansion results in two dimensions.
\end{abstract}

\maketitle

\section{I. Introduction}

Most of the fascinating phenomena in nature emerge from the collective behavior of microscopic elements that composed of the system. In general, these elements can be electrons in condensed matter systems, magnetic moments or spins in magnetic systems, repeated amino acids in proteins, genes in a cell, or even neurons that involved in our memories. The interactions among these constituent elements or units are responsible for these collective behavior. For example, collective behavior of many electrons in matter is responsible for superconductivity and magnetism, many amino acids determine the structure of a single protein, many genes determines the fate of a cell, and many neurons shape our thoughts. In most cases, the emergence of these collective behavior appear as a macroscopic order or sudden change in macroscopic properties in these systems. In these many repeated unit systems, statistical mechanics can provide the microscopic description of the system. In general, the connection between microscopic statistical mechanics description and macroscopic thermodynamics is provided by so called partition function,

\begin{eqnarray}
Z = Tr[e^{-\beta H}] \\ \nonumber
Z = \sum_i e^{-\beta E_i}
\end{eqnarray}

\noindent where inverse dimensionless temperature $\beta = 1/k_BT$ and $H$ is the Hamiltonian or the "energy function" of the system. In quantum mechanical sense, $E_i$'s are the eigenenergies of the Hamiltonian operator $H$.

Proteins are one of the fundamental building blocks of life and they present in almost all biological and cellular processes. Proteins consist of amino acids held together in a long chain by peptide bonds~\cite{u1, u2, u3, u4}. A given protein family generally has a similar amino acid sequence, three dimensional structures, and functions. Here the phrase "similar amino acids" refers to the same or same type of interacting amino acids. The evolutionary related amino acid sequence in a protein guides the protein folding process and its functional structure. When proteins function in biological processes, they \emph{fold} in to three-dimensional structures by curling the chain. The main theme of the protein folding problem is the question of how a protein's amino acid sequence dictates its three dimensional functional atomic structure~\cite{Folding}. Therefore, understanding the statistical patterns of amino acids sequence is very important when extracting structural information of a protein family. The difficulty of accurately predicting the structures from amino acid sequence is due to the fact that there are lot more sequences (more than 1000 times) than structures.

In order to effectively compare the variations of amino acids in a protein, a statistical approach known as multiple sequence alignment (MSA) of sequence data is used by matching up the chain position where the amino acids are often identical. The main idea of assembling sequence in this manner is to probe the statistical dependence of the data. This is done by an application of the maximum entropy principle in statistical mechanics to derive the distribution sequence or the determining the probability of appearing a certain amino acid sequence in a MSA. This approach is identical to the Gibbs distribution in statistical mechanics. The Gibbs distribution is extracted from an effective Hamiltonian that involve single site amino acid frequency and pairwise amino acid correlations. This effective Hamiltonian method is called "inverse problem" as one has to find the model parameters from large amount of observable data~\cite{reviewIN}.

In this work we introduce an effective Hamiltonian method in statistical mechanics to study the protein folding process. Here, we study the folding of proteins where adjacent repeat units pack against their neighbors resembling a lattice structure. Due to the interaction between repeated units, these molecular lattice of amino acids reach to thermal equilibrium showing emergence cooperative behavior to form the stable native structures.

In this paper, we investigate the thermodynamic properties of the protein folding process by modeling the protein using a $q$-state Potts model on a lattice~\cite{potts}. There are many studies on thermodynamics of protein, however different theoretical approaches have been used in previous studies~\cite{tr1, tr2, tr3, tr4, tr5, tr6, tr7, tr8, tr9, tr10, tr11}. These older approaches include, energy functional methods through contact probabilities~\cite{efcp1, efcp2, efcp3, efcp4}, variational approaches within functional integrals~\cite{vpfi1, vpfi2, vpfi3}, Ising-like model approaches~\cite{ism1, ism2}, beyond-Ising-like models~\cite{bism1, bism2, bism3, bism4}, combinational approaches~\cite{cmb1}, and computational methods~\cite{com1, com2, com3, com4, com5}. In this paper, we treat the protein folding process from a statistical perspective where the folding transition take places as a competition of entropy and energy (see FIG.~\ref{mod} below). First, we consider the each amino acid in the protein as a single unit siting in a lattice structure, and use a variational mean-field theory to convert our interacting model into an effectively non-interacting model. Then by investigating the free energy functional as a function of folding order parameter, we find that the protein folding is a strongly first order phase transition. Second, we investigate the effect of local environment through a uniform external field for a nearest-neighbor one dimensional system using transfer matrix method. We find that the correlation length calculated within this exact treatment shows a peak indicating the on-set of protein folding in the presence of external field. Third, we use a large-$q$ expansion to the $q$-state Potts model in two dimensions as a test of the accuracy of our mean field result. For two- dimensional nearest-neighbor model, we find that the mean-field theory over estimates the folding temperature by $9\%$.

The paper is organized as follows. In section II, we introduce the $q$-state Potts model and discuss its connection to the protein folding. In section III, we use a variational mean-field approach to the Potts model to investigate the protein folding transition and provide our mean-field results. In section IV, we solve the one-dimensional protein system exactly using a transfer matrix method and discuss its results with respect to the protein folding. We devote section V to introduce a large $q$-expansion to the two dimensional $q$-state Potts model and compare the results with mean-field results. Finally in section VI, we summarize our results with a discussion.

\section{II. Modeling Proteins with the Potts Model}

\begin{figure}
\includegraphics[width=\columnwidth]{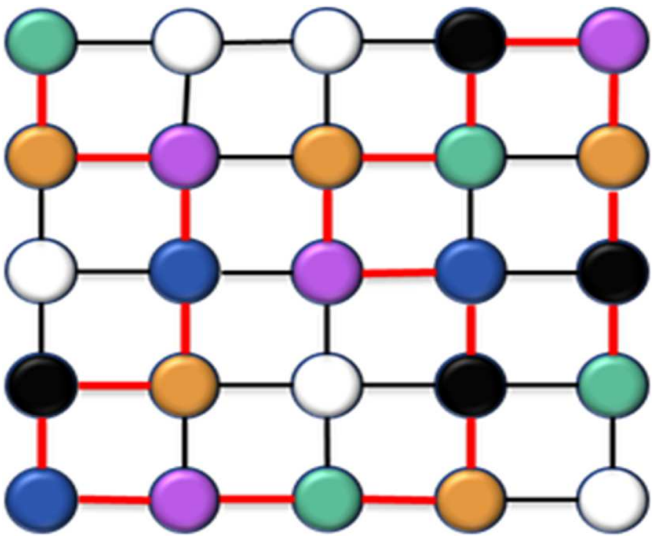}
\caption{(color online) The two-dimensional protein is represented as a chain of amino acids (colored dots) connected by peptide bonds (red lines). While different colored dots occupied at lattice sites represent the different amino acids, the white dots represent the empty sites on the two dimensional lattice. When the entropy dominates at higher temperatures, the chain stretches inside the lattice by moving amino acids to empty sites. When the energy dominates at lower temperatures, favored by the Potts model, same color amino acids cluster together by moving them closer. This clustering and stretching of same colors due to the competition between the energy and the entropy represents the protein folding and un-folding in our model.}\label{mod}
\end{figure}

A protein is a polypeptide chain consists of a sequence of amino acids. The sequence of these amino acid units is called primary structure. In this primary structure, the units or amino acids are connected by peptide bonds between the carbon atom of one unit and the nitrogen atom of the neighboring one. The center of each amino acid is a carbon called alpha-carbon. Therefore, the peptide chain of protein can be viewed as repeating peptide units that connect one alpha-carbon atom to another one along the backbone. All atoms in such a unit form a single plane, though neighboring units can be in a different plane. Not worrying about the detailed configuration of the side chains, the position of alpha-carbon atom along the chain is considered as different “sites” joined by the peptide bond as shown in FIG.~\ref{mod}. Therefore, we treat each amino acid in the protein as a single site corresponding to the alpha-carbon atom in the real protein~\cite{ac1, ac2, ac3, ac4, ac5}.

The folding of a peptide chain into a three dimensional structure is a thermodynamical driven process such that the chain naturally evolves to form domains of similar amino acids. The formation of this domain occurs by curling the one dimensional amino acid sequence by moving similar amino acids proximity to each other. Therefore, understanding the formation of this domain structure or the ordering of amino acids is crucial for predicting three-dimensional protein structures. Due to the lack of complete theories, qualitative explanations for many puzzling features of the kinetic and thermodynamics of biological self-organization process of proteins must be drawn from approximate models. The kinetic behavior of protein folding is more complicated than the thermodynamics behavior. In this work, we expect to illuminate the thermodynamical protein folding features from a statistical mechanics model known as the $q$-state Potts model. This model captures the attractive interaction between identical residues of the protein. However, it does not take into account the repulsive interaction between positively charged polar residues. As a result, our study will capture \emph{only} the qualitative features of the protein folding process. Within our model, the protein folding emerges as a collective aspect of attractive residues. The FIG.~\ref{mod} schematically describes the physics behind our modeling. For sophisticated realistic interactions in proteins, one needs to construct advanced theoretical models to capture the quantitative complex features in the folding process, as has been proposed recently~\cite{ac6, ac7}.

Here we study the ordering of amino acids assuming they are arranged in a lattice structure as discribed in FIG.~\ref{mod}. The folding mechanism of protein is very robust so that macroscopic thermodynamics properties do not depend on the microscopic details of the model. We study the macroscopic properties of folding process using $q$-state Potts model. First, we define a variable (spin) $\sigma_i$ at lattice site $"i"$ that takes $q$ discrete values, $\sigma_i = 1, 2, \cdot \cdot \cdot, q$ representing the amino acid sequence of a domain. The unique integer $q$ represents each domain of different amino acids. In this study we take $q$ to be equal $21$, representing $20$ naturally occurring amino acids and one additional state for gaps or empty spaces. The gap state must be available for an amino acid to move when they fold into three dimensional structures to form domains. Then the Hamiltonian representing interacting amino acids on a lattice is given by~\cite{pottsRe},

\begin{eqnarray}
H = -\frac{1}{2}\sum_{i \neq j} J_{ij} \delta_{\sigma_i, \sigma_j} -\sum_ih_i \delta_{\sigma_i, \sigma_{q_0}},
\end{eqnarray}

\noindent where
\begin{eqnarray}
\delta_{\sigma_i, \sigma_j} = \left\{
                                \begin{array}{ll}
                                  1, & \hbox{if $\sigma_i = \sigma_j$;} \\
                                  0, & \hbox{if $\sigma_i \neq \sigma_j$.}
                                \end{array}
                              \right.
\end{eqnarray}

\noindent This is the well-known $q$-state Potts model in an external field $h_i$ favoring variables to align in $\sigma_{q_0}$; that is forming domains of amino acid $\sigma_{q_0}$. The model allows each lattice site to have one out of $q$ different states. Here $J_{ij}$ is the pairwise interaction strength or the exchange parameter between variables $\sigma_i$ and $\sigma_j$ at two different lattice sites $i$ and $j$. The Kronecker delta function $\delta_{\sigma_i, \sigma_j}$ and the negative sign in front of the first term favor to have same amino acids at sites $i$ and $j$. Due to this pairwise attractive interaction between same residues, domains of same amino acids is expected to be formed by folding the amino acid chain into a three dimensional structure. Within the $q$-state Potts model, this structural formation is characterized by "ordering" of spin variables $\sigma$ in a specific state. In the following, we detect this "ordering" by a order parameter which indicates whether amino acids have domain formation or not. The magnitude of the order parameter measures the amount of ordering or the formation of similar amino acid domains. Even though the influence of solvent is not explicitly included in the Potts model, its effect is indirectly included in the model through interaction parameters and spin variables. For example, the local environment creates by the solvent, such as pH value, is controlled by the field $h$. The solvent-residue interactions, such as hydrophobic effects, are included through the variable $\sigma$ and pairwise interaction $J_{ij}$. As a result, all hydrophobic residues tend to group or form domains proximity to each other by ordering in a specific state $\sigma = \sigma_{hp}$, representing hydrophobic amino acids.

When $q = 2$, using the identity  $\delta_{\sigma_i, \sigma_j} = 1/2 (1 + \sigma_i \sigma_j)$ and assuming two distinct values of $\sigma_i = \pm 1$, this Potts model becomes the well known Ising model in an external field if we set $J_{ij} \rightarrow 2 J_{ij}$. Then the 2-state long-ranged Ising model has the form~\cite{Ising},

\begin{eqnarray}
H = -\sum_{i \neq j} J_{ij} \sigma_i \sigma_j -\sum_ih_i \sigma_i.
\end{eqnarray}

\noindent If $J_{ij} = J_0$, independent of neighbors and only consider the neighboring pairs, this is called nearest-neighbor Ising model (NNIM) in an external field. NNIM is a well known and simplest statistical model that has been studied by Ising to explain the magnetic phase transition of magnetic materials. Although, Ising managed to solve the model exactly in one dimension, the model does not show any phase transitions. The absence of phase transition is general for any discrete nearest-neighbor spin models in one-dimension, including Potts model introduced above.

\section{III. Variational Mean Field Theory for the $q$-state Potts model}

As we mentioned before, the general investigation of  protein folding problem requires to solve the inverse Potts model, \emph{i.e.} finding the coupling constants  $J_{ij}$ and $h_i$ from the data base. In this study we take those to have certain structures and study the Potts model within the protein folding context. We treat exchange interaction $J_{ij}$ to have two different forms in one-dimension, $J_{ij} = J_\alpha |r_{ij}|^{-\alpha}$ and  $J_{ij} = J_\lambda e^{-|r_{ij}|/\lambda}$. The first one is the long-range interaction representing the algebraic decay with the inter unit distance and the second one represents short, intermediate, and extended interactions for $\lambda < a$, $a < \lambda < 2 a$, and $2 a < \lambda$, respectively. Here $r_{ij} = r_i - r_j$ is the distance between two amino acids (or gaps) on the lattice and it is a multiple of the lattice constant $a$. For three-dimensional and two-dimensional lattices, we assume only nearest neighbors interaction with $J_{ij} = J_0$. For the variational mean-field theory introduced in this section, the effect of all types of interactions mentioned above can be incorporated into a single effective interaction parameter $J_z$ (defined later).

For the long-range $q$-state Potts model has four parameters, the temperature ($T$), external field ($h$), the number of states $q$, and decay parameters $\alpha$ and $\lambda$, thus the model is expected to show rich behavior. Even though the physical basis remain unclear, the long-range interactions are very common in proteins. As approximate lattice models of proteins have proven to capture some of the basic properties of real proteins, we believe our study will elucidate some general principles of protein stability and folding.

In order to study the thermodynamics properties of protein folding process, we wish to derive Helmholtz free energy $F = U-TS$, of the system, where $U$ is the internal energy and $S$ is the entropy with $T$ being the temperature. In terms of density matrix $\rho$, which is related to the partition function as $\rho = e^{-H/k_BT}/Z$, the internal energy and entropy can be written as $U = Tr[\rho H]$ and $S = -k_B Tr[\rho \ln \rho]$, respectively, where $k_B$ is the Boltzman constant and the $Tr[A] = \sum_{i = 1}^N \sum_{\sigma_i =1}^q A$ is the trace of a matrix $A$. The density matrix $\rho = \prod_{i =1}^N \rho_i(\sigma_i)$ is a product of individual density matrices at each site $i$ which are functions of spin variables $\sigma_i$. For our variational mean field theory, we take our variational density matrix at site $i$ as $\rho_i = (1-m_i)/q + m_i \delta_{\sigma_i, \sigma_{q_0}}$, where the local order parameter $m_i$ is corresponding to the ordering of spin $\sigma_i$ in state $\sigma_{q_0}$ and can be written as,

\begin{eqnarray}
m_i = \frac{q \langle \delta_{\sigma_i, \sigma_{q_0}} \rangle - 1}{q - 1}.
\end{eqnarray}

\noindent Here the expectation value of $\delta_{\sigma_i, \sigma_{q_0}}$ is defined as $\langle \delta_{\sigma_i, \sigma_{q_0}} \rangle = Tr [\delta_{\sigma_i, \sigma_{q_0}} e^{-\beta H}]/Z$. Observing the behavior of this expectation value with the Hamiltonian $H$ in the high- and low-temperature limits; $ \displaystyle \lim_{T \to \infty} \langle \delta_{\sigma_i, \sigma_{q_0}} \rangle = 1/q $ and $\displaystyle \lim_{T \to 0} \langle \delta_{\sigma_i, \sigma_{q_0}} \rangle = 1$, the local order parameter gives $ \displaystyle \lim_{T \to \infty} m_i = 0$ and $\displaystyle \lim_{T \to 0} m_i = 1$ representing the disordered and perfect ordered phases, respectively. Therefore, the order parameter is in the range $ 0 \leq m_i \leq 1$. The change of $m$ value from zero to a finite value as one decreases the temperature indicates the thermal phase transition from an entropy dominated disordered phase to an energy dominated ordered phase. While entropy dominated phase represents the unfolded phase, the energy dominatd phase represents the protein folded phase. Notice that our density matrix is normalized as $Tr[\rho] = \prod_{i = 1}^N Tr_i[\rho_i] = \prod_{i = 1}^N \sum_{\sigma_i = 1, 2, ....q} \rho_i(\sigma_i) = 1$. Using our variational density matrix, we find the internal energy ($U$) and the entropy ($S$),

\begin{eqnarray}
U = -\frac{1}{2q} \sum_{i \neq j} J_{ij}[1 + (q-1) m_i m_j] \\ \nonumber - \frac{1}{q}\sum_{i =1}^N h_i[1 + (q-1) m_i]
\end{eqnarray}

\begin{eqnarray}
S = -\frac{k_B}{q}\sum_{i =1}^N \biggr\{\biggr((1-q)(m_i-1) \ln\biggr[\frac{1-m_i}{q}\biggr] \\ \nonumber + \biggr([1+(q-1)m_i] \ln\large[\frac{1 + (q-1)m_i}{q}\biggr]\biggr\}.
\end{eqnarray}

\noindent We assume external field is uniform along the peptide chain so we set $h _i = h$ so the order parameter $m _i = m$ for all sites. In the thermodynamic limit where $N \rightarrow \infty$, we find the free energy per site $f(m) = F(m)/N$ as,

\begin{eqnarray}
f(m) = -\biggr(\frac{1 + (q-1) m^2}{2 q}\biggr) J_{z} - \biggr(\frac{1 + (q-1) m}{q}\biggr)h \\ \nonumber
 + \frac{k_BT}{q}\biggr\{(1-q)(m-1)\ln\biggr(\frac{1-m}{q}\biggr) \\ \nonumber + [1 + (q-1)m] \ln \biggr[\frac{1 + (q-1)m}{q}\biggr]\biggr\},
\end{eqnarray}

\noindent where $J_{z} = z\sum_{i \neq j} J_{ij}$ represents the effective exchange interaction with coordination number $z$. The coordination number $z$ is simply the number of nearest neighbors. For a three-dimensional cubic lattice, a two-dimensional square lattice, and a one-dimensional lattice, $z$ has the values $6$, $4$, and $2$, respectively. For the case of one-dimensional long-ranged and exponential pairwise interaction strengths $J_{ij} = J_\alpha |r_{ij}|^{-\alpha}$ and  $J_{ij} = J_\lambda e^{-|r_{ij}|/\lambda}$, we have $J_{z} = 2z J_\alpha a^{-\alpha} \sum_{n=1}^\infty  \frac{1}{n^\alpha} \equiv 2 J_\alpha a^{-\alpha} \zeta (\alpha)$  and $J_{z} = 2z J_\lambda \sum_{n=1}^\infty e^{-n a/\lambda} \equiv  2J_\lambda \frac{1}{e^{a/\lambda}-1}$, respectively, where $\zeta (\alpha)$ is the Riemann zeta function. The free energy of the nearest-neighbor only interaction spins can be obtained by simply replacing $J_{z}\rightarrow zJ_0$, where $J_0$ is the nearest-neighbor interaction parameter. The order parameter $m$ is determined by minimization of the free energy, $df/dm =0$. This condition gives us two solutions for the order parameter $m$, $m =0$ and the second one is given by the solution of a self-consistent equation,

\begin{eqnarray}
m = \frac{e^{\eta}-1}{e^{\eta} +q -1}
\end{eqnarray}

\noindent where $\eta = (h + m J_{z})/(k_BT)$. We find the critical temperature $T_c$ by assuming order parameter is small close to the phase transition and then expanding the free energy in powers of $m$ as a Landau energy functional,

\begin{eqnarray}
f(m) = f_0 + c m + \frac{A}{2} m^2 -\frac{y}{3} m^3 + \frac{D}{4} m^4,
\end{eqnarray}

\noindent where $f_0$, $c$, $A$,$y$, and $D$ are all temperature dependent constants. Notice that $D = (k_B T/3) (-3 + 6  q - 4  q^2 +  q^3)$ is always positive for $q \geq 2$, so the Landau energy functional is bounded from below as required for the stability. For non-zero values of $c = h(1/q - 1) < 0$, the minimum of Landau energy functional is at $m > 0$ for all temperatures, thus the system is at the ordered phase at any finite temperature for non-zero $h$ values. In the presence of cubic term $ y =  (k_B T/2) (2  - 3  q +  q^2)$ at $h =0$, the order parameter changes from zero to a finite value with a discontinuity at the critical temperature as one decreases the temperature, thus the phase transition is first order or discontinuous in nature unless $q = 2$. For $q =2$, the cubic term vanishes and the phase transition is determined by setting the coefficient of quadratic term $A = (1/q)(J_{z} - J_{z} q - k_BT q + k_BT q^2) = 0$. For this case, the order parameter continuously change from zero to a finite value at the critical temperature $k_BT_c = J_{z}/2$ as the temperature is lowered, hence the phase transition is second order or continuous. When the cubic term is present, the condition for critical temperature is derived by setting the two minima equal, $f(m = m_+) = f(m =0)$, where $m_+$ is the value of order parameter at the second minimum in free energy which is gained from the condition $df/dm = 0$. The condition for the critical temperature for first order phase transition is then given by $y^2 = (9/2) AD$. This condition gives us the critical temperature $k_BT_c = (6J_z/q) (3-3q+q^2)/(14-14 q + 5q^2)\rightarrow 0.056 J_z$ for $q =21$. For strongly first order transitions, the order parameter $m$ at the critical temperature is not a small parameter. As a result, the series expansion of $m$ given in Eq. (10) is not valid and the critical temperature is most often under estimated. Instead, we can work directly from the free energy and calculate the order parameter $m$ and other observable numerically.

\begin{figure}
\includegraphics[width=\columnwidth]{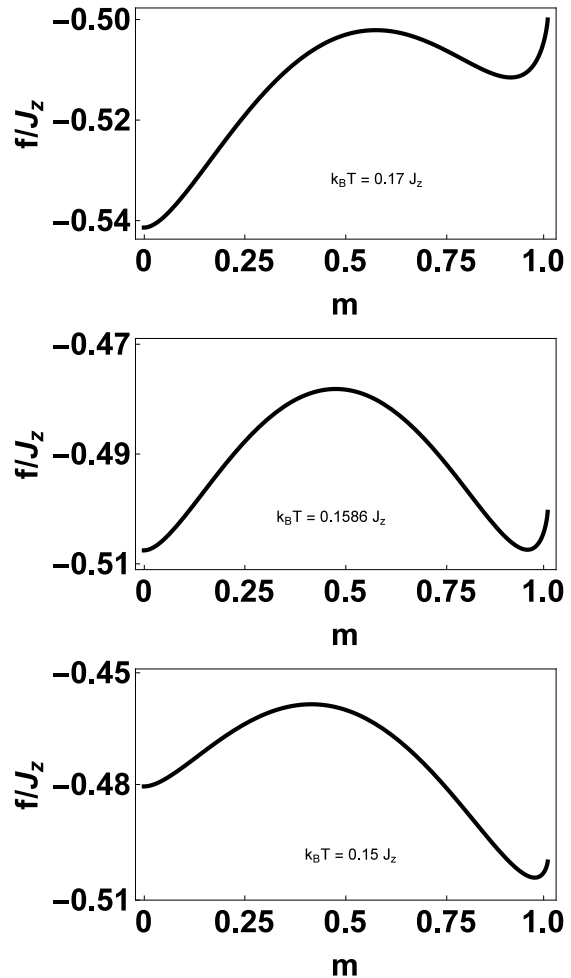}
\caption{The free energy profile for $q=21$ Potts model without external field ($h = 0$). The higher temperature local second minimum at $m \neq 0$ becomes the most minimum below the critical temperature $k_BT_C = 0.1586 J_z$. }\label{FE}
\end{figure}

\noindent For a first order phase transition, the solutions of self-consistent Eq. (9) does not guarantee the identification of the critical temperature. For first order phase transitions, the transition occurs when the second minimum in free energy profile becomes more favorable than that of the first minimum at $m =0$. The evolution of the free energy profile is shown in FIG.~\ref{FE}. At higher temperatures, a second minimum develops at non-zero values of $m$, yet the most minimum is at $m =0$. As the temperature is lowered, the second minimum coincides with the first minimum at a critical temperature showing the first order phase transition. Below this critical temperature $k_BT_C = 0.1586 J_z$, the second minimum remains as a global minimum representing the ordered state.

\begin{figure}
\includegraphics[width=\columnwidth]{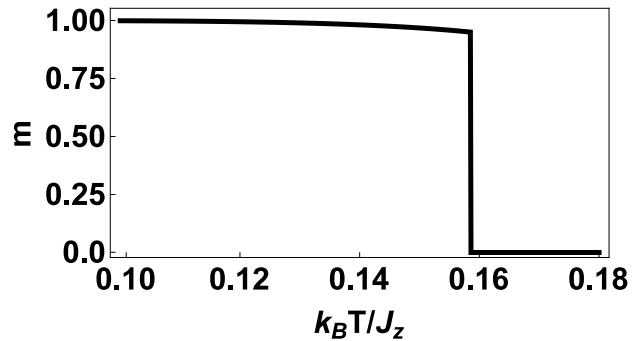}
\caption{The temperature dependence of the protein folding order parameter for $q=21$ Potts model without external field ($h = 0$).}\label{op}
\end{figure}

\noindent After finding the critical temperature through the free energy profiles, we solve the self-consistent Eq. (9) numerically to calculate the order parameter $m$. The order parameter as a function of temperature is shown in Fig.~\ref{op}. Notice that the temperature is scaled with effective coupling constant $J_z$, therefore the results shown in FIG.~\ref{FE} and FIG.~\ref{op} are valid for one, two, and three dimensional systems. As seen from the FIG~.\ref{op}, the order parameter is zero at higher temperatures and gets a finite value below the critical temperature $k_BT_C = 0.1586 J_z$ representing the protein folding. For the nearest-neighbor only interaction model in one dimension, critical temperature can be written in terms of nearest-neighbor only interaction parameter $J_0$ by setting $z =2$ and  we find $k_BT_C/J_0 = 0.32$. As has been already known and evidence by the next section, the nearest-neighbor only interaction model in one dimension does not show any finite temperature ordering. However, the Potts model with long range interactions exhibit long range order at finite temperatures for certain values of $\alpha$~\cite{1dct}. In higher dimensions, we expect that the mean-field theory to provide reasonable estimate to the critical temperatures even for nearest-neighbor only models. For a two dimensional square lattice and a three dimensional cubic lattice, the mean field critical temperatures are $k_BT_C/J_0 = 0.63$ and $k_BT_C/J_0 = 0.95$, respectively. These critical temperatures are reasonably comparable with scaling laws predicted within topology based simulations~\cite{ct1, ct2, ct3}.

\begin{figure}
\includegraphics[width=\columnwidth]{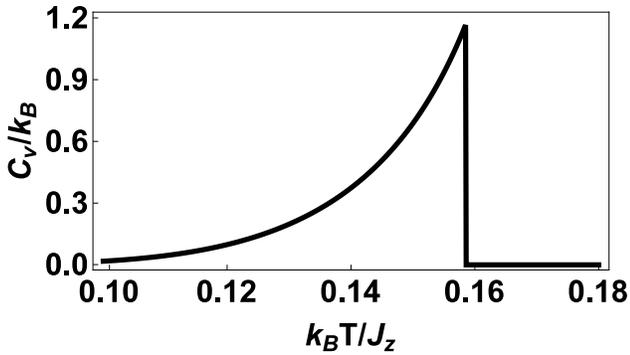}
\caption{The temperature dependence of the specific heat $C_v$  for $q=21$ Potts model without external field $h$.}\label{sph}
\end{figure}

\noindent Treating the temperature dependence of order parameter and then using $C_v = -T (\partial^2 f/\partial T^2)$, we calculate the specific heat $C_v$. The temperature dependence of the specific heat $C_v$ is also shown in FIG.~\ref{sph}. As can be seen from the figure, the specific heat shows a sharp peak at the protein folding transition temperature, thus, measurement of specific heat can be used as a detection of the protein folding in vitro.

\begin{figure}
\includegraphics[width=\columnwidth]{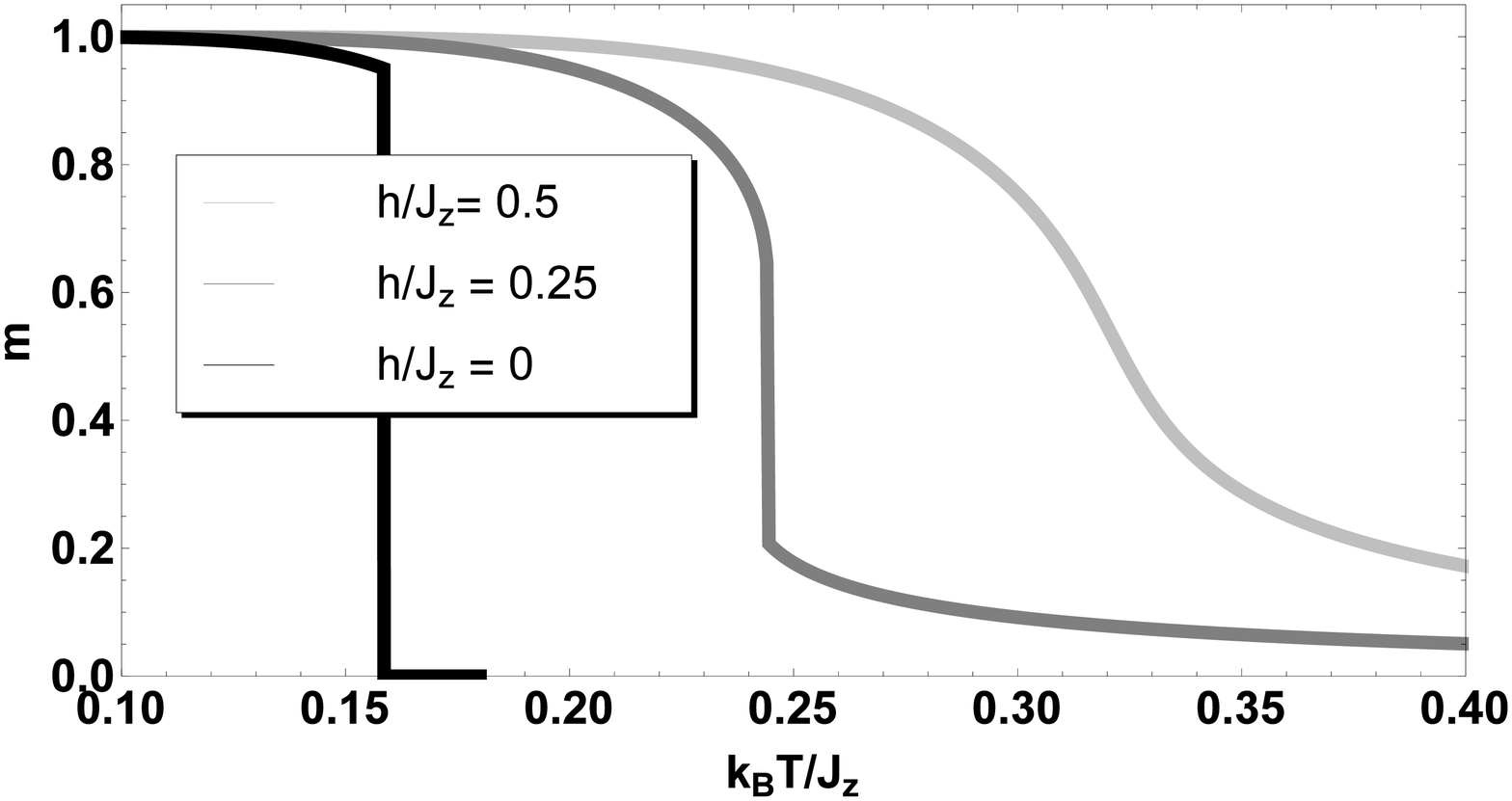}
\caption{The temperature dependence of the protein folding order parameter for $q=21$ Potts model for various values of uniform external field $h$. }\label{oph}
\end{figure}

The external field dependence on the order parameter as a function of temperature is given in FIG.~\ref{oph}. Notice that the temperature and external field are scaled with effective interaction parameter $J_z$ which depend on the number of nearest neighbors. As a result, the result shown in Fig.~\ref{oph} is also valid for all dimensions, including one dimension. The zero field non-zero order parameter in one dimension is due to the lack of accuracy in our mean-field theory. Thus, there is no phase transition in one dimension as we discuss in next section. Notice that for finite values of uniform external field $h$, the protein is always ordered as the order parameter is non-zero for any finite temperatures. However, for small values of $h$ values, one can see the sudden change in order parameter showing the vicinity of grouping most amino acids, perhaps showing a secondary transition.

\section{IV. Transfer matrix method for the one-dimensional $q$-state Potts model}

One-dimensional spin systems with periodic boundary conditions can be solved exactly~\cite{1dEX1, 1dEX2}. In this section, we derive the exact free energy of the $q$-state Potts chain using the transfer matrix method and compare it with our variational mean-field theory above. Here we assume that only the nearest neighbor interaction take place, however, the long-range interacting behavior can be obtained by replacing $J = J_0 \rightarrow J_{z}$ at the end of the derivation. The transfer matrix is a $q \times q$ matrix with $q$-number of eigenvalues~\cite{TRM}. Once the eigenvalues are known, the partition function can readily be calculated.

For a one-dimensional lattice with lattice sites $N$ and periodic boundary condition, \emph{i. e.} $\sigma_{N+1} = \sigma_1$, the partition function $Z = Tr[e^{-\beta H}]$ can be written as,

\begin{eqnarray}
Z = \sum_{\{\sigma_n\}} e^{\beta h \delta_{\sigma_1, \sigma_0}/2}e^{\beta J \delta_{\sigma_1, \sigma_2}}e^{\beta h \delta_{\sigma_2, \sigma_0}/2} \cdot \cdot \cdot \\ \nonumber
\cdot \cdot \cdot e^{\beta h \delta_{\sigma_N, \sigma_0}/2}e^{\beta J \delta_{\sigma_N, \sigma_1}}e^{\beta h \delta_{\sigma_1, \sigma_0}/2}.
\end{eqnarray}

\noindent This can be written as $Z = Tr[M^N]$, where $M$ is the $q \times q$ transfer matrix with elements $M_{\sigma, \sigma^\prime} = e^{\beta h \delta_{\sigma, \sigma_0}/2}e^{\beta J \delta_{\sigma, \sigma^\prime}}e^{\beta h \delta_{\sigma^\prime, \sigma_0}/2}$. The matrix elements in the transfer matrix are,

\begin{eqnarray}
M_{\sigma, \sigma^\prime} = \left\{
  \begin{array}{ll}
    e^{\beta (J+h)}, & \hbox{if $\sigma = \sigma^\prime = \sigma_0$;} \\
    e^{\beta J}, & \hbox{if $\sigma = \sigma^\prime \neq \sigma_0$;} \\
    e^{\beta h/2}, & \hbox{if $\sigma \neq \sigma^\prime$, $\sigma^\prime = \sigma_0$;} \\
    e^{\beta h/2}, & \hbox{if $\sigma \neq \sigma^\prime$, $\sigma = \sigma_0$;} \\
    1, & \hbox{ if $\sigma \neq \sigma^\prime \neq\sigma_0$.}
  \end{array}
\right.
\end{eqnarray}

\noindent The matrix $M$ has $q$-number of eigenvalues $\lambda_i$, with $i = 1, \cdot \cdot \cdot \cdot \cdot q$ and the partition function is then $Z = \sum_{i =1}^{q} \lambda_i^N$.  Using symbolic calculation in mathemtica, we find eigenvalues of the transfer matrix and find two of them are in the form $\lambda_1 = a + b$ and $\lambda_2 = a - b$, where $ a = (q-2 + e^{\beta J} + e^{\beta (J +h)})/2$ and $2 b = \sqrt{(q-2)^2 + 2(q-2) A + 2 B  + C}$, with $A = e^{\beta J}(1 - e^{\beta h})$, $B = e^{\beta h} (2q- 2 - e^{2 \beta J})$, and $C = e^{2 \beta J}(1 + e^{2 \beta h})$. The other eigenvalues are $(q-2)$ times degenerate and they are given by $\lambda_3 = e^{\beta J} - 1$. The free energy is then given by $F = -k_BT \ln Z$. In the thermodynamic limit where $N \rightarrow \infty$, the maximum eigenvalue $\lambda_1$ dominates and the thermodynamics is determined by the free energy $F = -Nk_BT \ln \lambda_1$.

In the absence of external field $h$, the nearest-neighbor model considered in this section does not show any finite temperature phase transitions. As the mean-field theory is more accurate only for higher dimensions, the zero-field finite temperature protein folding transition for a purely one-dimensional model obtained in previous section is an overestimation of the mean-field theory. However in the presence of an external field $h$, even a purely one-dimensional model can show indication of folding transition. In order to investigate the effect of external or local field, we calculate the exact \emph{correlation length} ($\xi$) within our transfer matrix method. The correlation function is a measure of how the local fluctuations in one part of the chain affect those in another part. In other words, it is a measure of how amino acids in one part of the chain influence the amino acids in another part to undergo folding. Such influences occur over a characteristic distance known as the correlation length. The correlation length can be defined in terms of first two largest eigenvalues as,

\begin{eqnarray}
\frac{1}{\xi} = \ln [\lambda_1/\lambda_2].
\end{eqnarray}

\begin{figure}
\includegraphics[width=\columnwidth]{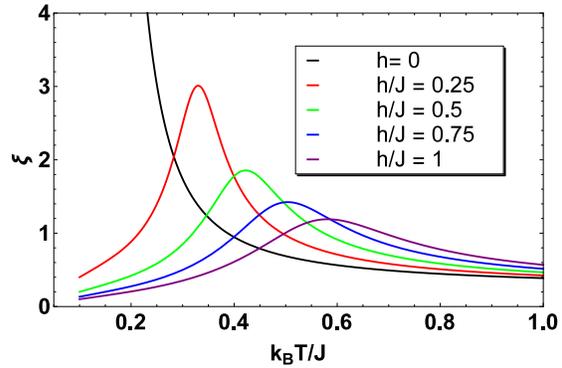}
\caption{(color online) The correlation length $\xi$ as a function of temperature calculated from transfer matrix method for the one dimensional $q=21$-state Potts model. The sharp peaks at non-zero external field $h$ indicates the onset of protein folding transition in one dimension.}\label{CL}
\end{figure}

\noindent The calculated correlation length ($\xi$) as function of temperature ($T$) is shown in FIG.~\ref{CL}. As seen from the figure, in $h \rightarrow 0$ limit, the correlation length diverges at $T = 0$. This zero temperature divergence indicates the absence of protein folding phase transition at any finite temperatures for nearest neighbor model in a zero field. This contrast to the finite temperature folding transition obtained from the mean-field theory in previous section. Notice the finite temperature sharp peaks in correlation length at non-zero field $h$. These represent the onset of protein folding for the one-dimensional Potts model in the presence of an external field.

\section{V. Large $q$-limit of the two-dimensional $q$-state Potts model}

In the absence of exact solutions, approximation methods such as variational mean-field theory discussed in Sec. III are used to investigate the properties of model systems. If the system Hamiltonian possesses a small or a large parameter, one can expand the partition function in terms of the small parameter or the inverse of large parameter and then safely neglect the higher order terms. The high-temperature and the low-temperature expansion analysis remain as one of the useful tools in many -particle systems~\cite{series}.

In this section, we combine the high-temperature and the low-temperature series expansion for the two-dimensional square lattice Potts model to develop an effective large $q$ series expansion to the partition function~\cite{LqE}. As the $q$-value for our system is relatively large, one can derive an accurate partition function by truncating the higher order terms as we discuss below. For simplicity, we restrict ourselves to the nearest-neighbor $q$-state Potts model on a square lattice without external field $h$. Then the partition function can be written as;

\begin{eqnarray}
Z = \sum_{\sigma_i}\prod_{\langle ij \rangle}e^{K\delta_{\sigma_i, \sigma_j}},
\end{eqnarray}

\noindent where $K = \beta J_0$. Writing $e^{K\delta_{\sigma_i, \sigma_j}} = U_K[1 + W_K (q \delta_{\sigma_i, \sigma_j}-1)]$ with the high-temperature expansion parameter $W_K$ and evaluating this for $\sigma_i =\sigma_j$ and  $\sigma_i \neq \sigma_j$ cases separately to solve for $U_K$ and $W_K$, one finds~\cite{pottsRe},

\begin{eqnarray}
W_K = \frac{e^K-1}{e^K + q -1} \nonumber \\
U_K = \frac{e^K + q -1}{q}.
\end{eqnarray}

\noindent Then using the $N$ number of squares of 4-bonds for a square lattice, the high-temperature expansion of the partition function is given by,

\begin{eqnarray}
Z_H = U_K^{2N} q^{N}[1 + NW_K^4(q-1) + \cdot \cdot \cdot].
\end{eqnarray}

\noindent On the other hand, the minimum energy state is having all spins being in on of the $q$ possible states at low temperatures. The first excited state is having \emph{only} one site in a different spin state. The degeneracy factor for this case is $N \times (q-1)$ and the resulting energy cost is $4K$ due to the bonds around the different spin states. The resulting low-temperature partition function is then,

\begin{eqnarray}
Z_L = q e^{2NK}[1 + N(q-1)e^{-4K} + \cdot \cdot \cdot].
\end{eqnarray}

\noindent Defining a new variable $\tilde{K}$ through the duality relation,
\begin{eqnarray}
e^{-\tilde{K}} = \frac{e^K-1}{e^K+q-1},
\end{eqnarray}

\noindent one can combine the low-temperature free energy per site $f_L(K) = -\ln Z_L/(N\beta)$ and the high-temperature free energy per site $f_H(K) = -\ln Z_H/(N\beta)$~\cite{LqE},

\begin{eqnarray}
f_H(K) = -2 \ln \biggr[\frac{e^K-1}{\sqrt{q}}\biggr] + f_L(\tilde{K}).
\end{eqnarray}

\noindent Finally, defining a small parameter $v = (e^K-1)/\sqrt{q}$ for large values of $q$ and then using the duality condition, we combine the high-temperature and the low-temperature free energies to get a series expansion of the free energy for the system,

\begin{eqnarray}
f = f_H(v) + \Theta(v-1)[f_H(1/v) - f_H(v) + 2 \ln v],
\end{eqnarray}

\noindent where $\Theta(v)$ is the Heaviside step or unit step function and the function $f_H(v)$ can be evaluated up to the seventh order in $v$ with first two terms in partition functions $Z_H$ and $Z_L$,

\begin{eqnarray}
f_H(v) = -2 \ln \sqrt{q} - \frac{2}{\sqrt{q}} v + \frac{1}{q} v^2 -\frac{2}{3q^{3/2}}v^3 + \frac{3-2q}{2q^2}v^4 \\ \nonumber + \frac{20q-22}{5q^{5/2}} v^5 + \frac{31-30q}{3q^3} v^6 + \frac{140q-142}{7q^{7/2}} v^7 + \cdot \cdot \cdot.
\end{eqnarray}

\noindent Replacing $f_H(v)$ in Eq. (20) with this series expansion completes the large-$q$ free energy construction of the $q$-state Potts model. The investigation of the constructed free energy shows a discontinuity at $v =1$. This discontinuity indicates the first order protein folding phase transition at a critical temperature $k_BT_C = 0.58 J_0$. By comparing this large-$q$ expansion result with the mean-field result discussed in Sec. III, we find that the mean field theory over estimates the critical temperature by $9\%$ for the nearest-neighbor two dimensional square lattice.

\section{VI. Discussion and conclusions}

The heart of the protein folding problem is understanding the question of how a protein's amino acid sequence dictates its structure. The various interaction parameters inside the protein and the local environment are directly responsible for the folding. However in this research, we coded all these effects inside the pairwise interaction parameter and the Potts variables in our statistical model. This simplifying treatment allowed us to discuss the protein folding process from a perspective of statistical physics where the folding of protein resulted due to the competition between the entropy and the energy of the system.

In conclusion, we have studied the thermodynamics of protein folding process from a statistical perspective. We modeled the protein using $q$-state Potts model where each amino acids is treated as a single unit that can sit on a lattice. We then tackle the interacting Hamiltonian using a variational mean-field theory. We found entropy dominated unfolded state of the protein undergoes a strongly first order transition into an energy dominated folded state as one decreases the temperature. In one and two dimensions, we tested our mean-field results by using a transfer matrix method and a large-$q$ expansion method, respectively. In one dimension, we found that the correlation length shows a sharp peak at the onset of protein folding transition in the presence of a small external field. Our investigation of the large-$q$ expansion method for the two dimensional square lattice indicated that the mean-field theory over estimates the critical temperature by $9\%$. In this study, we treated Only attractive interactions between identical residues of proteins and used cyclic boundary conditions. As a result our findings are qualitative, however our study shows the power of statistical mechanics approaches in understanding the biological systems.

\section{VII. Acknowledgments}

 The authors acknowledge the support of Augusta University, the medical college of Georgia's TRP travel award, and the CURS faculty-student award. TD acknowledges the hospitality of ITAMP at the Harvard-Smithsonian Center for Astrophysics and KITP at the UC-Santa Barbara. The KITP visit was supported in part by the National Science Foundation under Grant No. NSF PHY11-25915.


\begin{references}



\bibitem{u1} Barrick D., Ferreiro D.U., Komives E.A, Curr. Opin. Struct. Biol. \textbf{18}, 27 (2008).
\bibitem{u2} Kajander T., Cortajarena A.L., Main E.R., Mochrie S.G., Regan L. A, J. Am. Chem. Soc. \textbf{127}, 10188 (2005).
\bibitem{u3} Kloss E., Courtemanche N., Barrick D, Arch. Biochem. Biophys. \textbf{469}, 83( 2008).
\bibitem{u4} Mello C.C., Barrick D, Proc. Natl. Acad. Sci. USA. \textbf{101}, 14102 (2004).


\bibitem{Folding} Ken A. Dill, S. Banu Ozkan, M. Scott Shell, and Thomas R. Weikl, Annu. Rev. Biophys. \textbf{37} (2008).
\bibitem{reviewIN} For example, see the review by Thierry Mora and William Bialek, J. Stat Phys. \textbf{144}, 268 (2011).

\bibitem{potts} R. B. Potts, Proc. Camb. Phil. Soc. \textbf{48}, 106 (1952).

\bibitem{tr1} S. Plotkin and J. N. Onuchic, Quarterly Reviews of Biophysics \textbf{35}, 111 (2002).
\bibitem{tr2} L. Mirny and E. Shakhnovich, Annu. Rev. Biophys. Biomol. Struct. \textbf{30}, 361
(2001).
\bibitem{tr3} H. S. Chan and K. A. Dill, Proteins Struct. Funct. Genet. \textbf{30}, 2 (1998).
\bibitem{tr4} J. D. Bryngelson, J. N. Onuchic, N. D. Socci, and P. G. Wolynes, Proteins
Struct. Funct. Genet. \textbf{21}, 167 (1995).
\bibitem{tr5} J. N. Onuchic, H. Nymeyer, A. E. Garc´ıa, J. Chahine, and N. E. Socci, Adv.
Prot. Chem. \textbf{53}, 87 (2000).
\bibitem{tr6} H. S. Chan and K. A. Dill, Annu. Rev. Biophys. Chem. \textbf{20}, 447 (1991).
\bibitem{tr7} M. Oliveberg and P. Wolynes, Quarterly Reviews of Biophysics \textbf{38}, 245 (2005).
\bibitem{tr8} W. A. Eaton, V. Mu˜noz, S. J. Hagen, gouri S. Jas, L. J. Lapidus, E. R. Henry,
and J. Hofrichter, Annu. Rev. Biophys. Biomol. Struct. \textbf{29}, 327 (2000).
\bibitem{tr9} B. Gillespie and K. Plaxco, Annu. Rev. Biochem. \textbf{73}, 837 (2004).
\bibitem{tr10} C.D.Snow, E. SOrin, Y. Rhee, and V. Pande, Annu. Rev. Biophys. Biomol.
Struct. \textbf{34}, 43 (2005).
\bibitem{tr11} M. Gruebele, Annu. Rev. Phys. Chem. \textbf{50}, 485 (1999).

\bibitem{efcp1} B. A. Shoemaker, J. Wang, and P. G. Wolynes, Proc. Natl. Acad. Sci. USA \textbf{94},
777 (1997).
\bibitem{efcp2} B. A. Shoemaker, J. Wang, and P. G. Wolynes, J. Mol. Biol. \textbf{287}, 675 (1999).
\bibitem{efcp3} B. A. Shoemaker and P. G. Wolynes, J. Mol. Biol. \textbf{287}, 657 (1999).
\bibitem{efcp4} S. Plotkin and J. N. Onuchic, Proc. Natl. Acad. Sci. USA \textbf{97}, 6509 (2000).
\bibitem{vpfi1} J. J. Portman, S. Takada, and P. G. Wolynes, Phys. Rev. Lett. \textbf{81}, 5237 (1998).
\bibitem{vpfi2} J. J. Portman, S. Takada, and P. G. Wolynes, J. Chem. Phys. \textbf{114}, 5069 (2001).
\bibitem{vpfi3} J. J. Portman, S. Takada, and P. G. Wolynes, J. Chem. Phys. \textbf{114}, 5082 (2001).
\bibitem{ism1} R. Zwanzig, A. Szabo, and B. Bagchi, PNAS \textbf{89}, 20 (1992).
\bibitem{ism2} R. Zwanzig, PNAS \textbf{92}, 9801 (1995).
\bibitem{bism1} V. V. Mu˜noz,, P. Thompson, J. Hofrichter, and W. A. Eaton, Nature \textbf{390}, 196
(1997).
\bibitem{bism2} V. Mu˜noz, E. Henry, J. Hofrichter, and W. A. Eaton, Proc. Natl. Acad. Sci.
USA \textbf{95}, 5872 (1998).
\bibitem{bism3} P. Thompson, V. Mu˜noz, G. Jas, E. Henry, W. A. Eaton, and J. Hofrichter, J.
Phys. Chem. \textbf{104}, 378 (2000).
\bibitem{bism4} V. Mu˜noz and W. A. Eaton, Proc. Natl. Acad. Sci. USA \textbf{96}, 11311 (1999).
\bibitem{cmb1} E. R. Henry and W. A. Eaton, Chem. Phys. \textbf{307}, 163 (2004).

\bibitem{com1} R. Day and V. Daggett, Adv. protein chem, \textbf{66}, 37 (2003).
\bibitem{com2} H. A. Scheraga, M. Khalili, and A. Liwo, Annu. rev. phys. chem, \textbf{58}, 57 (2007).
\bibitem{com3} J. A. McCammon, B. R. Gelin, and M. Karplus, Nature, \textbf{267}, 585 (1977).
\bibitem{com4} M. Levitt, J. Mol. Biol, 168, 595 (1983)
\bibitem{com5} W. D. Cornell, P. Cieplak, C. I. Bayly, I. R. Gould, K. M. Merz, D. M. Ferguson, D. C. Spellmeyer, T. Fox, J. W. Caldwell, P. A. Kollman, J. Am. Chem. Soc, 117 (19),5179 (1995).


\bibitem{ac1} H. Taketomi, Y. Ueda, and N. Go, Int J Pept Protein Res \textbf{7}, 445 (1975).
\bibitem{ac2} C. Clementi, H. Nymeyer, and J. N. Onuchic, J. Mol. Biol. \textbf{298}, 937 (2000).
\bibitem{ac3} J.-E. Shea, J. N. Onuchic, and C. L. Brooks III, Proc. Natl. Acad. Sci. USA
\textbf{96}, 12512 (1999).
\bibitem{ac4} N. Koga and S. Takada, J. Mol. Biol. \textbf{313}, 171 (2001).
\bibitem{ac5} H.Kaya and H.S.Chan, J. Mol. Biol. \textbf{326}, 911 (2003).
\bibitem{ac6} E. Shakhnovich, Chem. Rev., 2006, 106 (5), pp 1559–1588.
\bibitem{ac7} X. Peng, A. K. Sieradzan, and A. J. Niemi, Phys. Rev. E \textbf{94}, 062405 (2016) and reference there in.
\bibitem{pottsRe} For example, see the tutorial review, F. Y. Wu, Rev. Mod. Phys. \textbf{54}, 235 (1982)
\bibitem{1dct} M. Aizenman, J. T. Chayes, L. Chayes, C. M. Newman, J Stat Phys (1988) 50: 1.
\bibitem{Ising} C. Domb, in \emph{Phase transitions and critical phenomena}, edited by C. Domb and M. S. Green (Academic London), Vol.3, p. 1, (1973).

\bibitem{ct1} N. Koga and S. Takada, J. Mol. Biol. \textbf{313}, 171 (2001).
\bibitem{ct2} P. G. Wolynes, Proc. Natl. Acad. Sci. USA \textbf{94}, 6170 (1997).
\bibitem{ct3} A. V. Finkelstein and A. Y. Badretdinov, Fold. Des. \textbf{2}, 115 (1997).

\bibitem{1dEX1} R. J. Baxter, \emph{Exactly Solved Models in Statistical mechanics}, Academic Press, 1982.
\bibitem{1dEX2} P. Martin, Potts Models and related Problems in Statistical mechanics, World Scientific, 1994.

\bibitem{TRM} B. Mirza, M.R. Bakhtiari, Physica A \textbf{343} (2004) 311–316

\bibitem{series} D. S. Gaunt and A. J. Gutttmann, in \emph{Phase transitions and critical phenomena}, edited by C. Domb and M. S. Green (Academic London), Vol.3, p. 181, (1973).
\bibitem{LqE} Ginsparg, Paul; Goldschmidt, Yadin Y.; Zuber, Jean-Bernard, Nuclear Physics, Section B, Volume 170, Issue 3, p. 409-432 (1980).

\end{references}
\end{document}